\documentclass[12pt]{article}
\usepackage{graphicx}
\usepackage{epsf}
\usepackage{latexsym}
\newcommand{\be}{\begin{equation}}
\newcommand{\ee}{\end{equation}}
\newcommand{\ba}{\begin{array}{l}}
\newcommand{\ea}{\end{array}}
\newcommand{\bqa}{\begin{eqnarray}}
\newcommand{\eqa}{\end{eqnarray}}

\begin{document}
\begin{center}

{\Large\sf The Constraints of Unitary on $\pi\pi$ Scattering
Dispersion Relations}
\\[10mm]
{\sc  Jianyang He, Z.~G.~ Xiao and H.~Q.~Zheng}
\\[5mm]
{Department of Physics, Peking University, Beijing 100871,
P.~R.~China }
\\[5mm]
\begin{abstract} A new dispersion relation for the partial
wave $\pi\pi$ scattering $S$ matrix is set up. Using the
dispersion relation we generalize the single channel unitarity
condition, $SS^+=1$, to the entire complex $s$ plane, which is
equivalent to the generalized unitarity condition in quantum
mechanics. The pole positions of the $\sigma$ resonance and the
$f_0(980)$ resonance are estimated based on the theoretical
relations we obtained. The central value of the $\sigma$ pole
position is $M_\sigma\simeq 410$MeV, $\Gamma_\sigma\simeq 550$MeV,
obtained after including the the constraint of the Adler zero
condition.

\end{abstract}
\end{center}

\vspace{1cm}
 In a series of recent
publications~\cite{xzsingle,xzcouple,ang}, the present authors
presented a dispersive approach  to discuss the single channel and
coupled channel $\pi\pi$ interaction physics. The essence of the
method is to use dispersion relations for physical quantities
containing poles and also cuts except those endowed by unitarity.
Different from more traditional methods, like the $K$ matrix
method, Pad\'e approximation, etc., in the present scheme  the
role of  the dynamical cuts can be traced explicitly. In
Ref.~\cite{xzsingle,xzcouple} we have established a dispersion
relation for $\sin(2\delta_\pi)$, where $\delta_\pi$ is the
$\pi\pi$ scattering phase shift defined in the single channel
unitarity region,
 \bqa
 \sin(2\delta_\pi)&=&\rho F \ ,\nonumber \\ F(s)&=&
 \alpha+ \sum_i{\beta_i \over 2i\rho(s_i)(s-s_i)}-\sum_j{1\over
  2i\rho(z^{II}_j)S'(z^{II}_j)(s-z^{II}_j)}  \nonumber \\ &&
  +{1\over\pi}\int_L{{\rm Im}_LF(s')
  \over s'-s} ds'+{1\over\pi}\int_R{{\rm Im}_RF(s')
  \over s'-s} ds'\ ,
   \label{sin2d}
  \eqa
where $F\equiv \frac{1}{2i\rho}(S-1/S)$ is the analytic
continuation to the real part of the scattering $T$ matrix defined
in the physical region (times a factor of 2). In
Eq.~(\ref{sin2d}), $s_i$ denote the possible bound state pole
positions and $\beta_i$ are the corresponding residues of $S$;
$z_j^{II}$ denote the possible resonance pole positions on the
second sheet. The integrals denote the cut contributions to
$\sin(2\delta_\pi)$, $L=(-\infty,0]$ is the left hand cut (LHC)
and $R$ starts from the $\bar KK$ threshold once the $4\pi$ cut
are neglected, $\alpha$ denotes the subtraction constant. The sum
of contributions from the cut integrals and from the subtraction
constant is sometimes called the background contribution.
 By evaluating the left hand integral in the
above dispersion relation using the $O(p^4)$ amplitude (and its
Pad\`e unitarization) of chiral perturbation theory result, it is
found that the LHC contribution to $\sin(2\delta_\pi)$ is negative
and concave in the I=J=0 channel. Therefore the $\sigma$ resonance
must be introduced to explain the experimental data.

The present note is a supplementary and an extension to our
previous studies. It will be shown that the discontinuities of the
partial wave $S$ matrix and all other physical quantities across
the unitarity cut can be expressed as an explicit dependence on
the kinematic factor $\rho=\sqrt{1-4m_\pi^2/s}$. In other words,
the presence of the discontinuity on the right is solely due to
the presence of the kinematic factor. Furthermore we are able to
re-express the unitarity constraint in a non-trivial and
$analytic$ expression which holds on the entire complex $s$ plane.
For the reason of simplicity we will confine our discussion in the
single channel unitarity region.  As an exercise we will estimate
the pole positions of the $\sigma$ and the $f_0(980)$ mesons using
the formalism discussed in this note. Some comments related to the
coupled channel physics will also be made.

We start from the single channel unitarity region, or more
precisely, the center of mass energy squared, $s$, is greater than
$4m_\pi^2$ and less than $16m_\pi^2$. The relation between the
partial wave unitary $S$ matrix and $T$ matrix is defined as \be
\label{STsingle}S(s)=1+ 2 i\rho(s) T(s), \ee
 where $\rho=\sqrt{1-4m_\pi^2 / s}\ .$
 With this definition the
single channel unitary relation,
 \be\label{uni}
 {\rm Im}T(s) =T(s) \rho(s) T^*(s)\ ,
 \ee
 is being used together with the property of real analyticity,
  \be
T^{I*}(s+i\epsilon)=T^I(s-i\epsilon)\ , \ee
 to  analytically
continue  the $S$ matrix and the $T$ matrix, which are analytic
functions on the physical cut plane, to the second sheet of the
Riemann surface:
 \be
T^{II}(s+i\epsilon)= T^I(s-i\epsilon)={ T^I(s) \over S^{I}(s)},
\,\, {\rm and}\,\,\, S^{II}= {1\over S^I}\ .\ee
 One can then
verify that the function $\tilde F$ defined as
 \be\tilde{F}\equiv
{1\over 2} (S+{1\over S})
 \ee
 has no discontinuity across the real axis when $0<s<16m_\pi^2$, since
 \bqa \tilde{F}(s-i\epsilon) &=&
\frac{1}{2}(S(s-i\epsilon)+{1\over S(s-i\epsilon)})
=\frac{1}{2}(S^{II}(s+i\epsilon)+{1\over S^{II}(s+i\epsilon)}) \nonumber \\
&=&\frac{1}{2} ({1\over S^I(s+i \epsilon)}+S^I(s+i\epsilon)
)=\tilde{F}(s+i\epsilon) , \eqa
 and
 the left hand cut it contains starts from $-\infty$ to $0$.
 The cut structure of $\tilde F$ is very similar to the cut structure of the
function $F$  studied previously.
 The function $\tilde{F}$ is
 the analytic continuation of $\cos (2 \delta_\pi)$ defined in the single
 channel unitarity region.

According to the analytic structure of $\tilde F$, as discussed
above we can set up the following dispersion relation,
 \bqa
  && \cos(2\delta_\pi) =\tilde{F}
   =\tilde{\alpha}+\sum_i {\beta_i\over2(s-s_i)}
  \nonumber\\
  &&+\sum_j{1\over
 2S'(z^{II}_j)(s-z^{II}_j)} +{1\over\pi}\int_L {{\rm
 Im}_L\tilde F(s')
 \over s'-s} ds' +{1\over\pi}\int_R {{\rm
 Im}_R\tilde F(s')
 \over s'-s} ds',  \label{cos2d}
\eqa
  where $\tilde\alpha$ is the subtraction constant and
   one subtraction to the cut integrals in the above
expression is understood. The right hand cut $R$ starts from
$16m_\pi^2$ in principle but becomes important only when $s$
approaches the $\bar KK$ threshold.~\footnote{We neglect the
$4\pi$ cut from now on in the text. According to the conventional
wisdom the $4\pi$ cut becomes important only above, say, 1.2GeV.}
Using Eqs.~(\ref{cos2d}) and (\ref{sin2d}), we get an analytic
expression of $S$ on the complex $s$ plane in terms of poles,
dynamical cuts, and the kinematic factor:
\bqa S(z)&=& \cos(2\delta_\pi)+i \sin(2\delta_\pi) \nonumber \\
&=&\tilde{\alpha}+i\alpha\rho(z)+\sum_i
{\beta_i\over2(z-s_i)}+\sum_i{\rho(z)\beta_i \over
2\rho(s_i)(z-s_i)}+\sum_j{\rho(z^{II}_j)-\rho(z)\over
  2\rho(z^{II}_j)S'(z^{II}_j)(z-z^{II}_j)}
  \nonumber\\&& +{1\over\pi}\int_L {{\rm Im}_L \tilde F\over s'-z} ds'
  +{i\rho(z)\over\pi}\int_L{{\rm Im}_L
 F\over s'-z} ds' +{1\over\pi}\int_R {{\rm Im}_R \tilde F\over s'-z} ds'
  +{i\rho(z)\over\pi}\int_R{{\rm Im}_R
 F\over s'-z} ds'. \nonumber\\
 \label{smatrix}
 \eqa
 One may use the definition $S(4m_\pi^2)=1$
 to re-express
  $\tilde \alpha$ in Eqs.~(\ref{smatrix}) and
 (\ref{cos2d}) in terms of other parameters. The
 above expression respects the well known properties of $S$ matrix
 theory. For example,
 the physical sheet $S(z)$ does not contain resonance poles
 though  the phase motion of $S$ is affected by resonance poles on the
  second sheet.  The Eq.~(\ref{smatrix}), though
simple to derive, is an exact relation.\footnote{Since all cuts at
higher energies are actually included in those cut integrals.} The
Eqs. (\ref{cos2d}) and (\ref{sin2d}) must satisfy a relation on
the whole complex $s$ plane:
 \be \sin^2 2\delta_\pi+\cos^2 2\delta_\pi \equiv 1,
 \label{constraint}
 \ee
 which is the analytic continuation of the single channel unitarity
 relation, $S^+S=1$, on the complex $s$ plane.
The Eq.~(\ref{constraint}) is equivalent to the generalized
unitarity condition $S(k)S(k^*)^*=1$ in quantum
mechanics~\footnote{We are in debt to the referee who points out
this fact to us.} and it contains all information about single
channel unitarity and analyticity. For example,
 Eq.~(\ref{smatrix}) must obey another relation,
  \be
\label{zero} S(z_j^{II})=0 .
 \ee
 This equation is derived by the analytic structure of $S$ matrix,
 that is, the physical $S$ matrix has zero at the same energy,
 $z_j^{II}$, as the pole energy on the second sheet.
 The Eq.~(\ref{zero}) is actually equivalent to the requirement of the vanishing
 of the first order pole terms on the $l.h.s.$ of
Eq.~(\ref{constraint}) (the second order poles disappear
automatically).  The Eq.~(\ref{constraint}) demands correlations
between various parameters: $s_i$, $\beta_i$, $z_j^{II}$,
$S'(z_j^{II})$, $\alpha$, and cut integrals. However, these
relations are in general very complicated to use directly.

We find it  helpful, for pedagogical reasons, to analyze
Eq.~(\ref{smatrix}) together with Eq.~(\ref{constraint}) in some
very simple situations. For example, we  neglect all the  cut
integrals in Eq.~(\ref{smatrix}), and assume only one pole at
$s=s_0$ exist. Then we found two solutions satisfying
Eqs.~(\ref{smatrix}) and (\ref{constraint}):
\begin{enumerate}
\item A bound state:
\begin{equation}
\tilde \alpha=1-s_0/2,\,\,\, \alpha=-{1\over
2}\sqrt{s_0(4-s_0)},\,\,\, \beta =s_0(4-s_0). \ee
 The scattering length is $a=-\sqrt{s_0\over 4-s_0}$  (taking the
 mass of the scattering particles to be 1).

\item A virtual state:
 \begin{equation} \tilde \alpha=1-s_0/2,\,\,\, \alpha={1\over
2}\sqrt{s_0(4-s_0)},\,\,\, \beta ={1\over S'(s_0)}=s_0(4-s_0). \ee
 The scattering length is $a=\sqrt{s_0\over 4-s_0}$.
\end{enumerate}
We can learn some  lessons from these two simple solutions.
Comparing with the nonrelativistic version of the toy
model~\cite{Barton}, \be\label{nonr} S={1+ika\over 1-ika}\ ,\ee
where $k=\sqrt{{s/ 4}-1}.$ The nonrelativistic version contains a
bound state pole  when $a<0$, and a virtual state pole  when
$a>0$. This agrees with the qualitative behavior of the
relativistic case. But the pole locates at $s_0=4(1-{1\over a^2})$
whereas in our case the pole locates at $s_0=4/(1+{1\over a^2})$.
In the nonrelativistic case the pole can locate at anywhere
between $-\infty$ and 4, but in the present case $s_0\in (0,4)$.
The latter of cousre makes sense by eliminating the possible
existence of tachyons. Furthermore, in the norelativistic case the
phase shift $\delta(\infty)$ goes to $\pm \pi/2$ as dictated by
the ``weak" Levinson's theorem. That is not the case in the
present situation, even there is no dynamical cut.\footnote{We
call the dynamical cuts as those appear as left hand integrals in
Eqs.~(\ref{cos2d}) and (\ref{sin2d}).} All these differences come
from the use of relativistic kinematics (to use $\rho$ instead of
$k$), which really makes physical sense, as shown above. The
relativistic kinematic factor introduces an additional cut from
the square root of $s$ which is conveniently placed at from 0 to
$-\infty$. This additional (kinematical) cut, though carefully
excluded from the dynamical cuts~\cite{xzsingle,xzcouple}, does
function in its own way. For solutions with more than one pole,
one can prove that a two--pole (a pair of resonances) solution
does not exist (in the absence of dynamical cuts), which is
different from the non-relativistic case. A three--pole solution
however exists. As an existence proof one can construct the $S$
matrix in the following form:
 \be
S=\frac{s-M^2-i\rho g}{s-M^2+i\rho g}\ ; \,\,\, g>0\ , \ee
 which, for $M^2>4$ and sufficiently small g, contains a pair of
 resonances and a virtual state pole. In general, however, there
 is no simple correspondance between $S$ matrix poles and the
 physical resonances~\cite{Beveren02}.

 In the phenomenological discussion on the realistic $\pi\pi$ scatterings,
 we
follow the method of Refs.~\cite{xzsingle,xzcouple} to study the
properties of resonance poles after the cut integrals are
estimated. In the following  we focus on the IJ=00 channel, using
the phase shift data from
Refs.~\cite{munich1,munich2,oldke4,newke4}. The difference between
the fit made in Refs.~\cite{xzsingle} and here is that in
Refs.~\cite{xzsingle} we only fit $\sin(2\delta_\pi)$, or the
imaginary part of the $S$ matrix. In here we fit the $S$ matrix
itself using Eq.~(\ref{smatrix}),
 \be
\label{logS1}
 \delta_\pi(s)={1\over 2i}\ln [S(s)]
 =
 {\rm Re}\left[{1\over 2i}\ln [S(s)]\right]+
 i\,{\rm Im}\left[{1\over 2i}\ln [S(s)]\right]\ .
 \ee
In the single channel unitarity region, $\delta_\pi$ is real. For
the given set of the experimental value of $\delta_\pi$: $\{s_j,
\delta_j,\Delta\delta_j\}$, one may construct the expression of
total $\chi^2$ containing two terms, $\chi^2=\chi^2_1+\chi^2_2$,
in which,
 \be\label{chi2}
\chi^2_1=\sum_j \frac{|\delta_j-{\rm Re}\left[{1\over 2i}\ln
[S(s_j)]\right]|^2}{|\Delta\delta_j|^2}\ ;\,\,\, \chi^2_2=\sum_j
\frac{|{\rm Im}\left[{1\over 2i}\ln
[S(s_j)]\right]|^2}{|\Delta\delta_j|^2}\ . \ee
 Notice that single
channel unitarity in here, unlike most conventional approaches, is
not guaranteed automatically. If we use the above expression of
$\chi^2$ to make the fit it may happen that the $\chi^2$
minimization program prefers a solution with non-vanishing
$\chi^2_2$, i.e., violating the single channel unitarity.
 In order to circumvent such a
problem the unitarity constraint Eq.~(\ref{constraint}) has to be
taken into account to confine the violation of unitarity in a
numerically acceptable range, which substantially complicates the
fit. What we gain with such a price paid is that we can, at least
in principle, clearly keep track of all kinds of dynamical
singularities in their right places. This property is not easy to
maintain in other approaches which automatically guarantee
unitarity.

 The circumvent is possible, noticing that the term
$\chi^2_2$ in Eq.~(\ref{chi2}) is in fact quite arbitrary, since
there is no experimental error bar for the `imaginary part of
$\delta_\pi$'. Therefore we can freely chose, for example, another
expression of $\chi^2_2$,
 \be\label{chi22}
\chi^2_2={1\over\epsilon^2}\sum_j |{\rm Im}\left[{1\over 2i}\ln
[S(s_j)]\right]|^2\ , \ee
 with sufficiently small $\epsilon$
parameter which will guarantee Eq.~(\ref{constraint}) in a
numerically satisfiable range. Actually what we do here is an
example of the so called `penalty function method' in the theory
of probability and statistics~\cite{zhu}. In here the term
$\chi^2_2$ defined in Eq.~(\ref{chi22}) is called the penalty term
and $1/\epsilon^2$ is called the penalty factor.

In order to make use of Eq.~(\ref{smatrix}) to study the
properties of resonance poles, it is necessary at first to
estimate various cut integrals.
 The discontinuities of function $F$ and
$\tilde F$ on the left can be rewritten as,
 \bqa {\rm Im_L}F&=&2{\rm Im_L Re_R}T(s)\ ,\label{F}  \\{\rm Im_L}
\tilde F&=&-2\rho(s){\rm Im_L Im_R}T(s)\ ,\label{tF} \eqa
 since $F=2{\rm Re}_R T$ and $\tilde{F}=1-2\rho {\rm Im}_RT$.
 The $r.h.s.$ of Eq.~(\ref{F}) has been estimated in
 Ref.~\cite{xzsingle},
 that is  one
expands ${\rm Im_L Re_R}T(s)$ to $O(p^4)$ in chiral perturbation
theory ($\chi$PT). But it is easy to see that ${\rm Im_L
Im_R}T(s)$ vanishes up to $O(p^4)$ since ${\rm
Im_R}T^{(4)}(s)=\rho\left({2s-m_\pi^2\over 32\pi F_{\pi}^2}
\right)^2$. Therefore ${\rm Im_L Im_R}T(s)$ must be expanded to
$O(p^6)$,
 \bqa
 {\rm Im_L Im_R}T(s) &=&2\rho T^{(2)} {\rm Im_L}{\rm
 Re}_RT^{(4)}(s)\ ,
 \eqa
 to get a non-vanishing result. Hence the bad high energy
 behavior of the chiral amplitude gets even worse when estimating
 Eq.~(\ref{tF}), which means when estimating the cut-off version
of the dispersion integral~\cite{xzsingle} the numerical result
will be very sensitive to the cut off parameter. In our
understanding the vanishing of ${\rm Im_L Im_R}T$ at  $O(p^4)$
implies that the quantity and its integral are indeed very small,
at least at moderately low energies. This suggestion is confirmed
by the prediction of the [1,1] Pad\'e amplitude which is very
small in magnitude.\footnote{Unlike the situation in the IJ=20
case where the spurious physical sheet resonance (SPSR)
contribution to $\cos (2\delta_\pi)$ is large~\cite{ang}, in the
IJ=00 case the SPSR contribution to $\cos (2\delta_\pi)$ is rather
small. The smallness of the SPSR contribution may be considered a
necessary condition for the predictions of Pad\'e amplitudes to be
numerically reasonable.} We therefore in the following fix the
left hand integral of $\tilde F$ by using the result from the
Pad\'e amplitude. We use the same strategy as in
Ref.~\cite{xzsingle} to estimate ${\rm Im}_LF$ and its integral.
That is we use both the $O(p^4)$ $\chi$PT  and the Pad\'e
approximant to estimate ${\rm Im}_LF$, for the former we truncate
the left hand integral at certain scale $-\Lambda^2$ which varies
within a reasonable range.

\begin{figure}[ht]
\mbox{\epsfxsize=6cm \epsffile{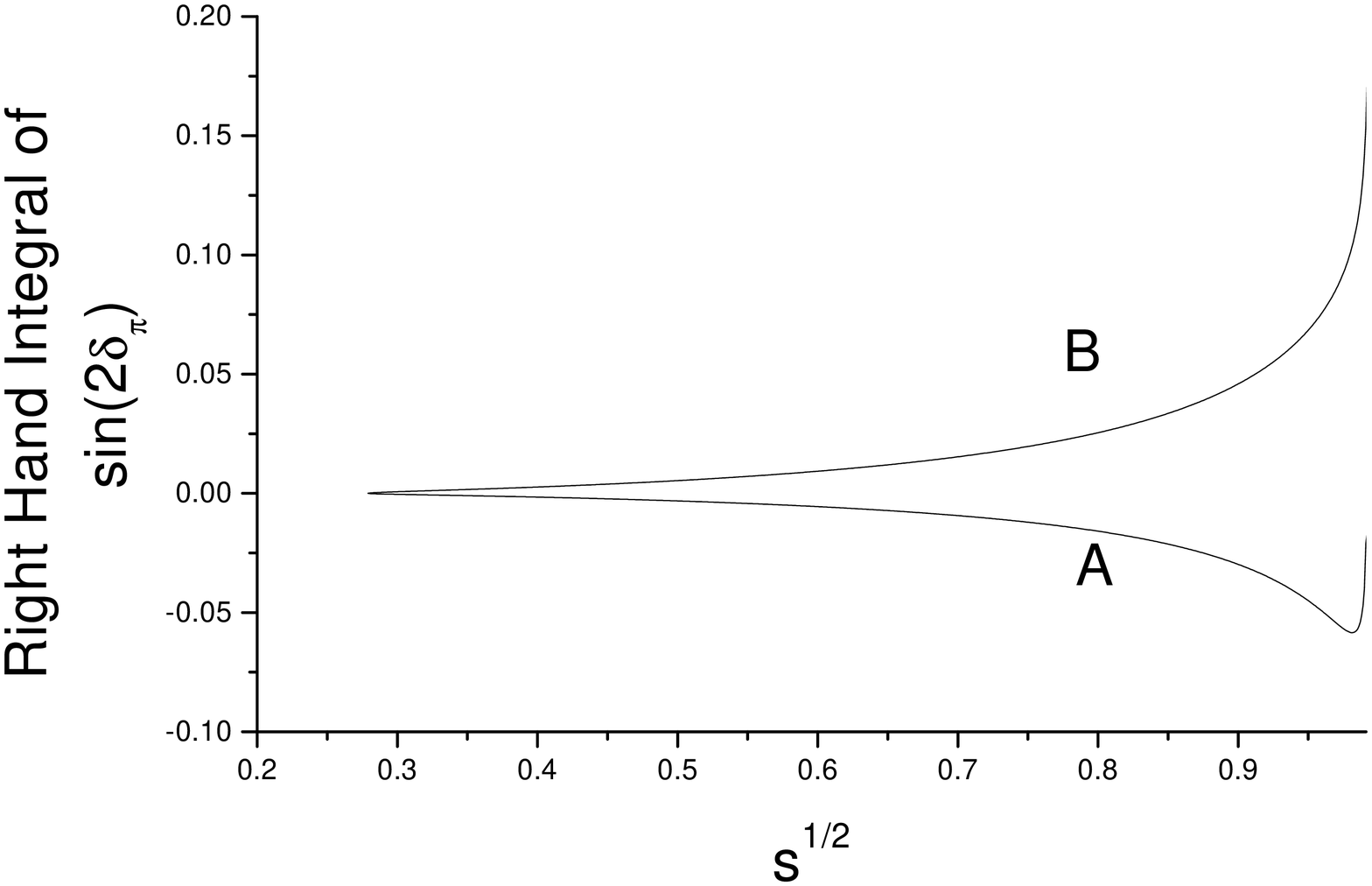}} \hspace*{1cm}
\mbox{\epsfxsize=6cm\epsffile{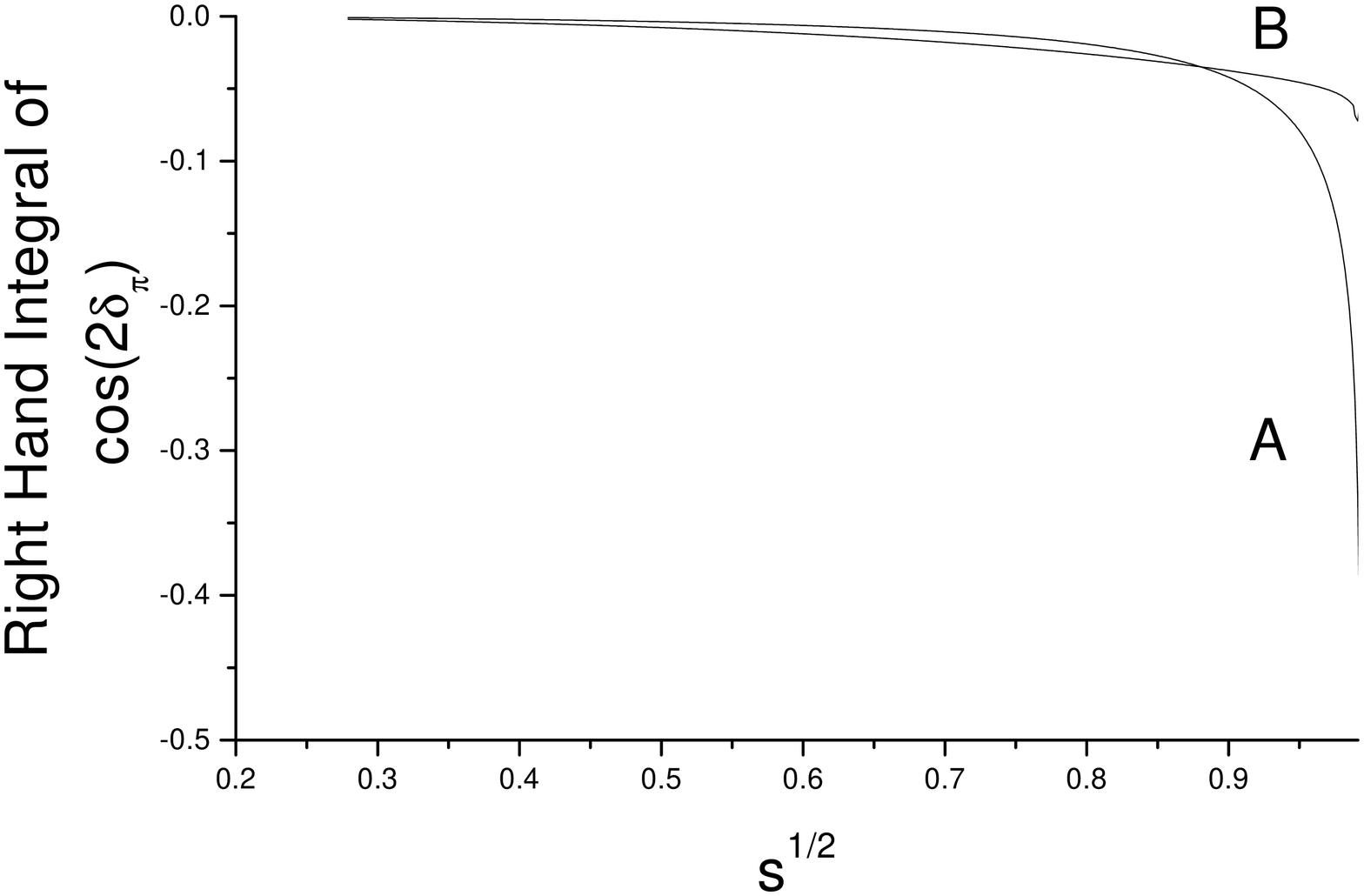}}
\caption{\label{rhcsincos} The contributions from right hand
integrals of $F$ and $\tilde{F}$. Line A is obtained from
Ref.~\cite{Zou93}, line B is obtained from Ref.~\cite{AMP}.}
\end{figure}

One of the lessons one may draw from Ref.~\cite{xzcouple} is that
it is not absolutely necessary to go to coupled channel situation
when discussing, at qualitative level, the property of the narrow
$f_0(980)$ resonance on the second sheet. Therefore we include the
$f_0$ pole in our discussion  within the current formalism which
only makes use of  the data in the single channel unitarity
region. In some sense, introducing the $f_0$ pole in the fit
improves the determination on the pole location of the $\sigma$
resonance as done in Ref.~\cite{xzsingle}, since in here we no
longer need to truncate the data (at around $\sqrt{s}\simeq
900$MeV) which is somewhat arbitrary. The right hand cut integrals
induced by the $\bar KK$ threshold have to be taken into account
in here since they will develop a cusp structure below the $\bar
KK$ threshold. Here we follow the same strategy as in
Ref.~\cite{xzcouple} to estimate the right hand integrals by using
the $T$ matrix parameterization above the $\bar KK$ threshold
given in Refs.~\cite{Zou93} and \cite{AMP}, and cut the integral
at $\sqrt{s}\simeq 1.5GeV$. For the integrand we have,
 \bqa {\rm Im_R}\tilde F&=&{1\over 2}(\eta -{1\over\eta})
 \sin(2\delta_\pi) , \nonumber
\\ {\rm Im_R}F
&=&{1\over 2\rho}({1\over\eta}-\eta)\cos(2\delta_\pi) . \eqa
  In fig.~\ref{rhcsincos},  we can see
the two estimates of the right hand integral in the IJ=00 channel.
Even though they are not coincide with each other,  both of them
give the same trend when approaching $4m_K^2$.
When we only fit $\sin(2\delta_\pi)$, the \textit{ r.h.c.} barely
have any effect to the pole position of $f_0(980)$. But in here,
we see that the effects of the right hand integrals  are no longer
negligible.

\begin{figure}[htb]
\begin{center}
\mbox{\epsfxsize=65mm \epsffile{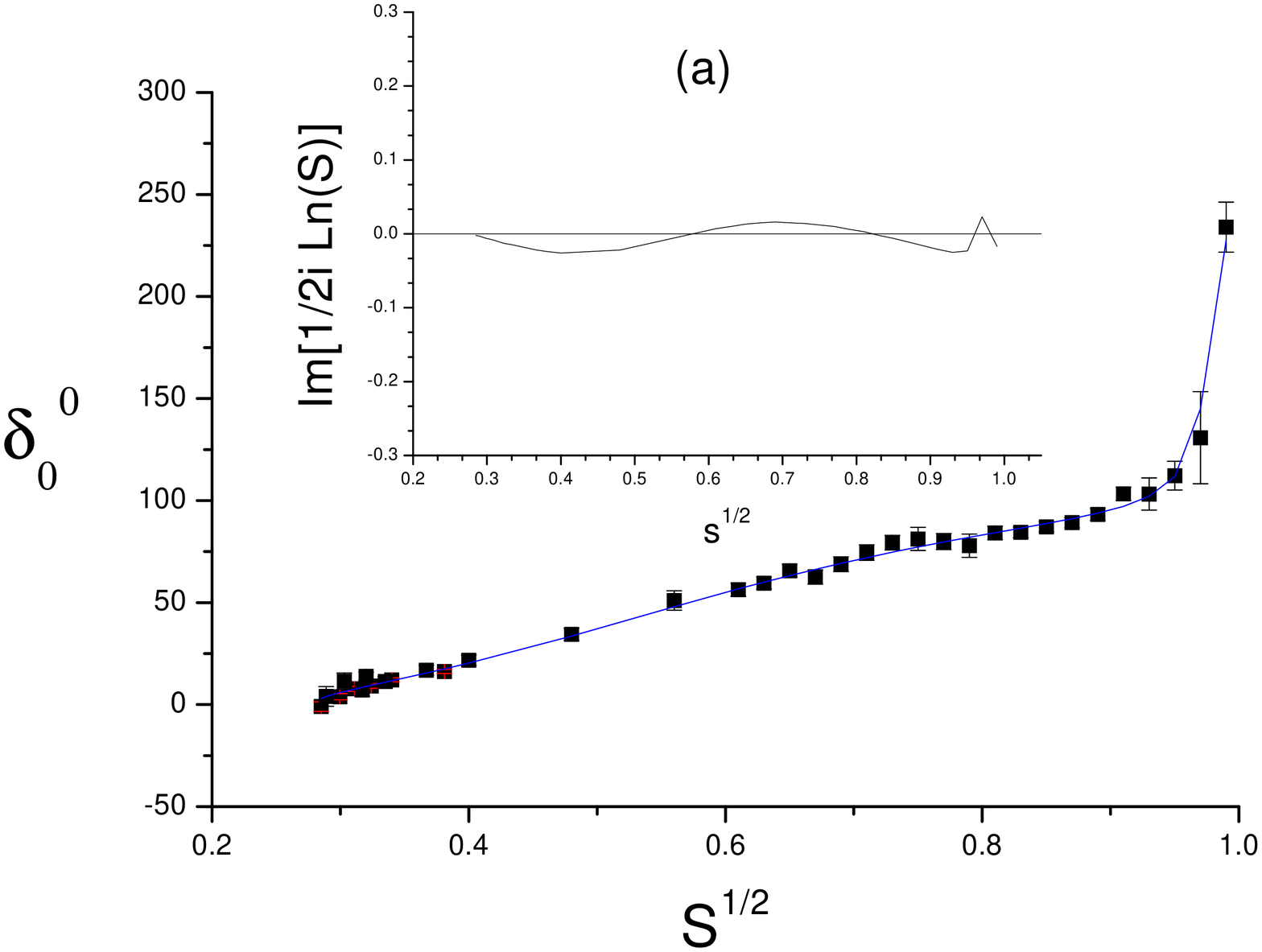}} \hspace*{10mm}
\mbox{\epsfxsize=65mm \epsffile{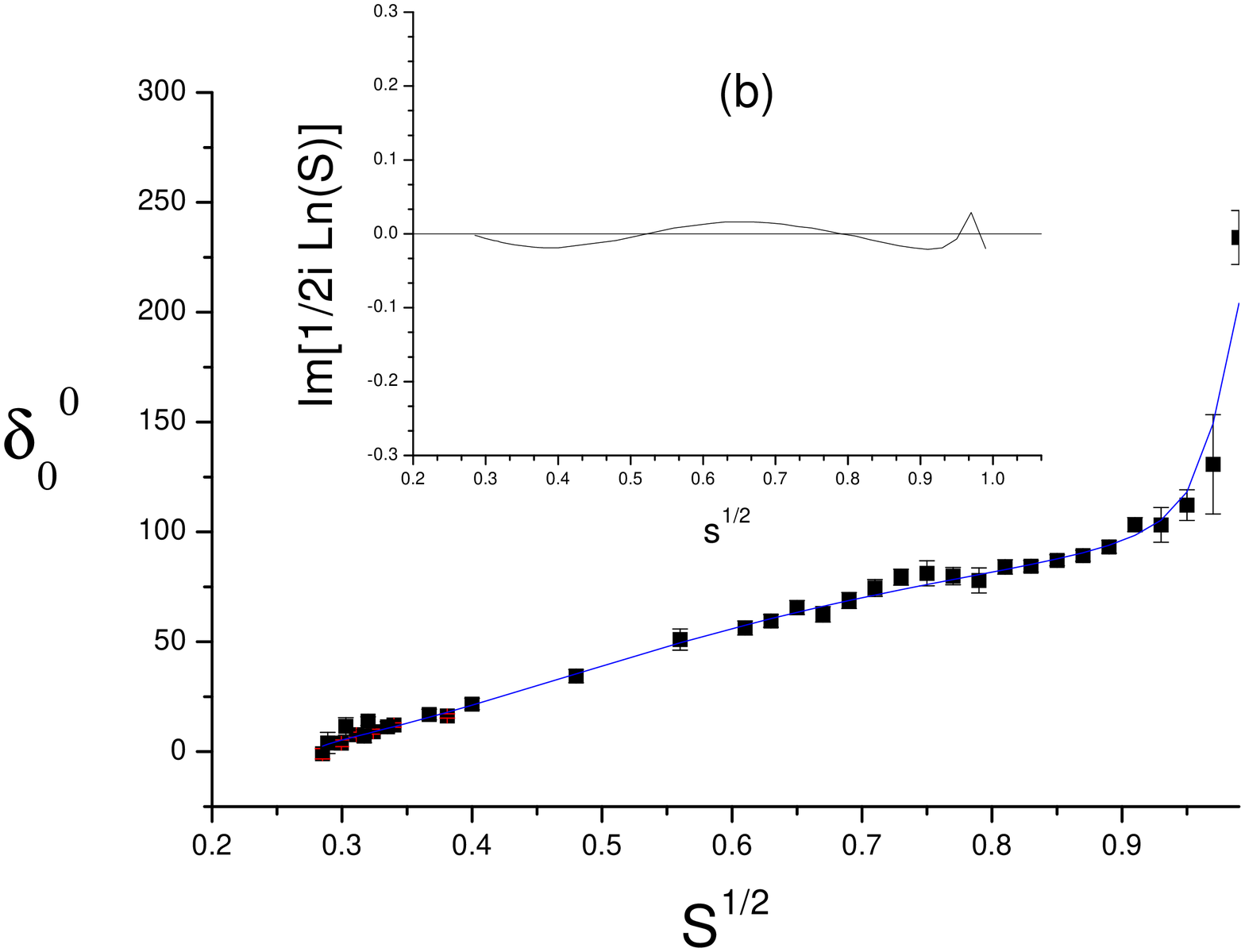}} \vspace*{-5mm}
\caption{\label{figlogS}(a): A fit using Eq.~(\ref{logS1}) and the
penalty function method. The $r.h.c.$ contributions are estimated
from Ref.~\cite{AMP}, the $l.h.c$ integrals are estimated from the
[1,1] Pad\'e amplitude and $\epsilon=0.02$. (b): The same as
Fig.~\ref{figlogS}a but with the constraint of Adler zero at
$s=m_\pi^2/2$.}
\end{center}
\end{figure}

As already stated earlier the unitarity  constraint,
Eq.(\ref{constraint}), has to be taken into account in our fit.
Instead of trying to solve the constraints among parameters
provided by Eq.(\ref{constraint}) explicitly we make use of the so
called `penalty function' method  in data fit with constraints
among parameters.  In principle, increasing the penalty factor
will  drive the fit result moving towards a solution respecting
unitarity exactly. But since there are uncertainties in the input,
i.e., the cut contributions and the number of pole terms,
increasing the penalty factor does not always lead to reasonable
results. For example, in the present case, for a too large penalty
factor corresponding to $\epsilon\sim 0.01$ the quality of the fit
near the $\bar KK$ threshold becomes very bad.\footnote{See
fig.~\ref{figlogS}, if one takes $\epsilon=0.01$ the fit curve of
$\delta_0^0$ would simply miss the data point which is just below
the $K\bar K$ threshold.}
The masses and widths of the $\sigma$ and the $f_0(980)$ poles can
be estimated from the  fit by varying the left and right cut
contributions. The variation range of the cutoff parameter
$\Lambda$  in  evaluating the left hand integral for $\sin
(2\delta_\pi)$ is taken from 600MeV to 800MeV here, as we find
that larger values of $\Lambda$  also lead the fit quality below
the $K\bar K$ threshold to be rather bad. The $\epsilon$ parameter
is therefore taken to be around 0.02. The results are listed in
the following:
 \bqa\label{woadler}
  && M_\sigma\simeq 440 - 530 {\rm MeV}\ ,\,\,\, \Gamma_\sigma\simeq 540 -
  590 {\rm MeV}\ ;\nonumber\\
 && M_{f_0}\simeq 976 - 987 {\rm MeV}\ ,\,\,\, \Gamma_{f_0}\simeq 22 -
  44 {\rm MeV}\ ;\nonumber\\
&&a_0^0\simeq 0.230 - 0.276\ .
  \eqa
  The above results
 are compatible with the results of Ref.~\cite{xzsingle} (the
 table~1 there), and especially
Ref.~\cite{xzcouple} (the Eq.~(44) there) though the methods are
somewhat different. The uncertainty for  the width of $f_0$ is
larger here when comparing with that of Ref.~\cite{xzcouple},
which may be partly due to the fact that in here we only work in
the single channel unitarity region. In Ref.~\cite{xzsingle}, the
unitarity constraint is not considered since only the imaginary
part of the $S$ matrix (or $\sin(2\delta_\pi)$ ) is fitted there.
When $\sin(2\delta_\pi)$ approaches 1, its error behaves as
$2\cos(2\delta_\pi)\Delta\delta_\pi$ and hence approaches 0.
Therefore the violation of unitarity is automatically confined in
an acceptable range in Ref.~\cite{xzsingle}.
 From the results we find
that the global fit favors a larger value of $a_0^0$ comparing
with the results of Refs.~\cite{newke4,CGL01}. A typical fit
result is plotted in fig.~\ref{figlogS}a.
 The problem of having a
larger scattering length  is due to that we have not put the
constraint of the Adler zero condition in our data fit. In
Ref.~\cite{xzsingle} this problem is solved by putting the
constraint of the scattering length parameter by hand.
Here we improve the fit by  including the constraint of the Adler
zero condition. If we fix the  Adler zero position at
$s=m_\pi^2/2$ the results are the following:
 \bqa\label{adler}
  && M_\sigma\simeq 380 - 440 {\rm MeV}\ ,\,\,\, \Gamma_\sigma\simeq 510 -
  580 {\rm MeV}\ ;\nonumber\\
 && M_{f_0}\simeq 976 - 983 {\rm MeV}\ ,\,\,\, \Gamma_{f_0}\simeq 43 -
  64 {\rm MeV}\ ;\nonumber\\
&&a_0^0\simeq 0.190 - 0.212\ .
  \eqa
The above results are obtained using the same range of parameters
as in obtaining Eq.~(\ref{woadler}). The most important change of
Eq.~(\ref{adler}) comparing with Eq.~(\ref{woadler}) is that now
the scattering length parameter decreases and is in better
agreement with the results of Ref.~\cite{newke4,CGL01}. We observe
again that the decreasing of the scattering length parameter
drives the $\sigma$ pole moving towards left on the complex $s$
plane, in agreement with the observation made in
Ref.~\cite{xzsingle}. Another major difference between
Eq.~(\ref{woadler}) and (\ref{adler}) is that the latter gives a
larger $f_0$ width. But Fig.~\ref{figlogS}b reveals that the $f_0$
pole is not fitted very well in the latter case. A more reliable
determination of the $f_0(980)$ resonance requires a coupled
channel analysis.

 In above discussions
 one of  the major uncertainty in obtaining our results comes
from the estimates on the left hand cuts which are very difficult
to determine accurately from pure theoretical calculations. Our
estimates on the $l.h.c.$ effects are based on $O(p^4)$ chiral
perturbation theory. It means that these estimates are no longer
trustworthy at high energies, or more precisely, at $s\simeq
m_\rho^2$ and above.
 Indeed, in our calculation to estimate ${\rm Im}F$
using $O(p^4)$ chiral perturbation theory, the resonance effects
are not taken into account because the resonance only contributes
at $O(p^4)$ the real part of the amplitudes. Taking the $\rho$
resonance as an example, the $t$ channel $\rho $ exchange will
enhance the t channel $\pi\pi$ cut at around $t=m_\rho^2$ and will
influence the $l.h.c.$ of $s$ channel partial wave amplitude
through Eq. (53) of Ref. [1]. This enhancement will result in a
contribution to the $l.h.c.$ starting from around
$4m_\pi^2-m_\rho^2$ to further left. If we approximate the effect
by a $tree$ $level$  $\rho$ exchange then the effect is
approximated by an effective $l.h.c.$ from $4m_\pi^2-m_\rho^2$ to
$-\infty$.\footnote{Notice that, exactly speaking, there is only
an enhancement to the $\pi\pi$ $l.h.c.$, but no new cut being
generated, since the $\rho$ resonance is not stable.} No
systematic method is known to exist to estimate these resonance
contributions to the dynamical cuts, in the non-perturbative
scheme.\footnote{A non-perturbative treatment is necessary because
otherwise the derivative coupling of $\rho$ to $\pi$ will cause
the once-subtracted dispersive integral divergent.} However, it is
reasonable to expect that the power counting rule for resonances
in perturbation theory~\cite{Ecker} also works in the
non-perturbation scheme, at low energies. Therefore it is a
reasonable speculation that resonance contribution to the left
hand integral is  $O(p^6)$ at low energies and therefore does not
distort any of our qualitative conclusion on the cut integral at
low energy, though their effects will become important at around
$s=m_\rho^2$ and above. Our speculation seems to be supported by a
phenomenological analysis of Ref.~\cite{ishida} (see fig.~2 in
that paper). A thorough investigation to this problem needs a
complete new study and goes beyond the scope of the present note.


 To conclude we in this note further extend the previous method
we proposed to study the partial wave  scattering problem by
establishing a dispersion relation for $\cos(2\delta_\pi)$. In our
procedure the effects of the unitarity cut are fully exposed by
the explicit dependence of physical quantities  on the kinematic
factor. The constraint of single channel unitarity,
Eq.~(\ref{uni}),  is re-expressed as an analytic relation which
holds on the whole $s$ plane, i.e., Eq.~(\ref{constraint}). The
Eq.~(\ref{constraint}) is equivalent to the generalized unitarity
condition in quantum mechanics. Applications of our approach are
made to determine the pole positions of the $\sigma $ and
$f_0(980)$ resonances, after estimating various cut contributions
from both chiral perturbation theory and experiments. We find that
the central value of the $\sigma$ pole position is about
$M_\sigma\simeq 410$MeV, $\Gamma_\sigma\simeq 550$MeV, according
to our fit with the constraint of the Adler zero condition.

One of the author (H.Z.) would like to thank Muneyuki Ishida for
helpful discussions. This work is supported in part by China
National Nature Science Foundation under grant number 10047003 and
10055003.

\end{document}